# Realisation of Highly Precise and Low Power Tunable Voltage Amplifier Based on the Translinear Circuit Scheme of CCCII+

**Umar Mohammad[1,2], , Mir Aamir Shafi[1]**

[1] Departement of Electronics and Communicating Engineering, Islamic University of Science & Technology, 192122
[2] Departement Electronics and Communicating Engineering, Integral University Lucknow, India.

**ABSTRACT**

In the past few years, advancements in the field of nano circuit design has become tougher than the demand. Low power devices have emerged tremendously.Both voltage mode aswell as current mode devices have proven alternative to each other for satisfying the demand of the growing market. As such, current conveyors have equitably established their uniqueness as an important circuit design element. The literature available to us during the few years in the field of analog VLSI design, quotes a huge number of application elements based on current conveyors. Likely, in this paper, a new tunable low power voltage amplifier based on the translinear circuit scheme of second generation current controlled current conveyor has been proposed. The modeling of the circuit presented in this paper employs the minimum number of passive elements. The magnitude of the tuning or the amplitude of the voltage presented here, is being controlled by means of two variable resistors. Current conveyor second generation translinear circuit scheme is taken into consideration to implement the proposed tunable voltage amplifier. CCCII works on the outlines of low power and low voltage design. Tunable voltage amplifiers find use in analog as well as in digital signal processing applications.

*Corresponding Author:*

**Umar Mohammad**
Department of Electronics and Communicating Engineering,
Islamic University of Science & Technology,
Awantipora, Pulwama , Jammu & Kashmir, India-192122
Author e-mail: ***umarnaik@iul.ac.in***

## 1. INTRODUCTION

Current conveyors **[1]** during the past few years have revolutionized the analog circuit design application area. Due to the low power and low voltage characteristics, the translinear circuit schemes of the current conveyors have proven worthy enough too. Secondly, the Current mode (CM) circuit design approach is firm attaining and establishing a trend setting reputation in the field of modern day VLSI. It proves to be a unique technique, which can help applying various design considerations which are ineffective or almost impossible to apply otherwise. In other terms its superiority over the voltage mode approach is trend setting **[2].**

Current Conveyor is a three terminal device, in which the input/output terminal are able to convey both current and voltage to each other. CCI, CCII, CCIII are the three generation of the current conveyors proposed by the Sedra, smith (CCI, CCII) **[3]** and Fabre (CCIII) **[4].** The use of CC's doesn't increase the complexity of the overall circuit designed. In this work, we are using the second generation current controlled current conveyor, because of the outstanding features like availability of band width, auto balancing, optimization of hardware, no triggering required etc.

A simple Voltage amplification scheme using CCCII+ is presented in this paper. The advantage of this type of amplifier is that, it requires a single CCCII+ circuit and two resistors, to amplify the input signal. The work presented in this paper has been Simulated and verified using HSPICE tool from Avant. The 45nm predictive modelling Parameters of Mosfet have been taken from the Arizona State University **[5].**

## 2. CCCII+ NON IDEALTIES AND DESIGN CONSIDERATIONS

The scheme, we have adopted in the second generation current conveyor is the current mirror based translinear circuit (CCCII+) **[6]** shown below in figure 2. The advantage of using this circuit scheme is that it is fully balanced. The intrinsic resistance in this type can be balanced by tuning the biasing current $I_b$.

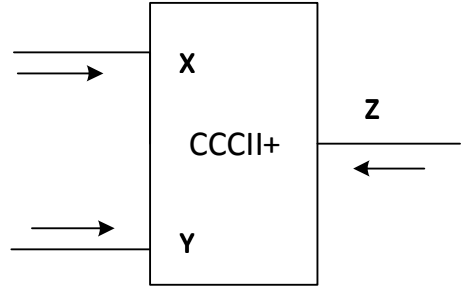

**Figure. 1:** General Representation of CCCII+

The matrix representation of CCCII is below;

$$\begin{bmatrix} V_X \\ I_y \\ I_z \end{bmatrix} = \begin{bmatrix} R_X & 1 & 0 \\ 0 & 0 & 0 \\ \pm 1 & 0 & 0 \end{bmatrix} \begin{bmatrix} I_X \\ V_Y \\ V_Z \end{bmatrix}$$

**Figure. 2:** Matrix Representation of CCCII± **[13]**

From the above matrix, we can conclude the following things relating the operation of CCCII±.

$$I_Y = 0 \qquad (1)$$

$$V_x = V_Y + I_X . R_X \qquad (2)$$

$$I_Z^+ = +I_x \qquad (3)$$

$$I_Z^- = -I_x \qquad (4)$$

**Resistance at X = $R_X$ = 1/2gm = 1/ $\sqrt{8.\beta_n.I_B}$,**

**Where $\beta_n = \mu_n C_{ox}.W/L$**

The above equations are well defined from the above matrix in figure 2, in which we have, "$I_Y$" is the node current at Port **"Y"**, "$V_x$" refers to voltage at node "**X**", $V_Y$ is the voltage at node **"Y"**. "$I_Z$" and "$V_Z$" are the current and Voltages at the node **"Z"** respectively. "$R_x$" is the intrinsic resistance of the translinear circuit.

Some of the promising features of the CCCII are **[7], [8]**;
  i. Greater Linearity
 ii. Availability of full bandwidth
iii. Autobalaing at Higher temperatures
 iv. Low Power, Low Voltage

## 3. FERRI, GIUSEPPE, AND NICOLA C. GUERRINI VOLTAGE AMPLIFIER USING CCII.

In literature, we have CCII, based voltage amplifier, proposed by Ferri, Giuseppe, and Nicola C. Guerrini **[9]**. In this circuit, there is a requirement of two CC's and two passive elements. The" Z" node in one of the CC's has been put in floating mode as shown in figure 3 below.

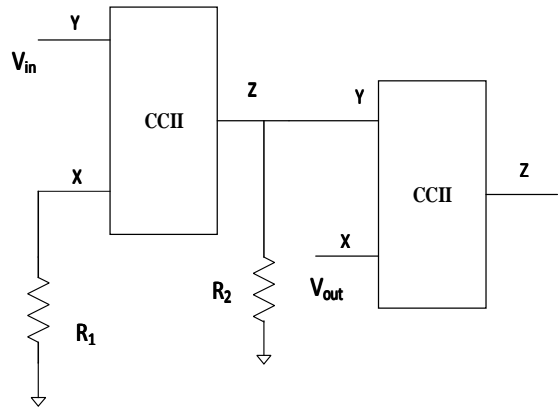

**Figure 3.** Ferri, Guerrini, CCII based Voltage Amplifier **[9].**

**4. PROPOSED TUNABLE VOLTAGE AMPLIFIER**
A variety of applications based on CCCII are available **[10] [11]**. But, as per the author's prior knowledge, voltage amplifier based on the scheme of CCCII+ is not available yet. However, a voltage amplifier employing one and two CCII's in available **[12].** The proposed circuit using CCCII+ performing Voltage amplification is provided below in figure 4, along with its voltage gain relationship. The advantage of using this type of current conveyor is the requirement of less hardware, low power consumption and less complexity.

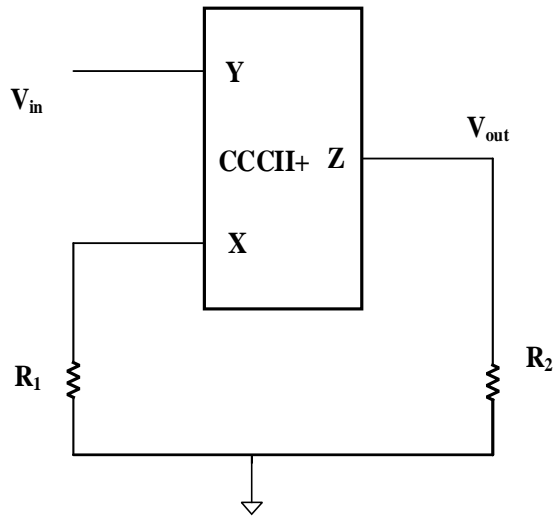

**Figure 4.** Proposed Tunable Voltage amplifier

The voltage gain relationship for the circuit proposed in figure 4, can be concluded as;

$$V_{out} = R_2 \cdot I_Z = R_2 \cdot I_X = R_2 \cdot \frac{V_X}{R_1}$$

$$V_{in} = V_Y = V_X$$

$$V_{out} = \frac{R_2}{R_1} V_{in}$$

The nomenclature of the circuit provided in figure 4 can be summarized as; At X, Y and Z are the three terminals of the CCCII+. $V_{in}$ is the input voltage at node X, $V_{out}$ is the output of the CCCII+ at node Z **[13],[14]**. $R_1$ and $R_2$ are the variable resistances provided at the nodes Y and X. The change in these two parasitic resistances are responsible for the increase and decrease in the amplitude of the input signal provided at node X. An increase in the value of the resistances provided, a gradual decrease in the amplitude of the Input signal is seen. Similarly lowering the values of the resistances, we can observe a sharp increase in the amplitude of the input signal. The translinear circuit scheme used in this Current conveyor is shown below;

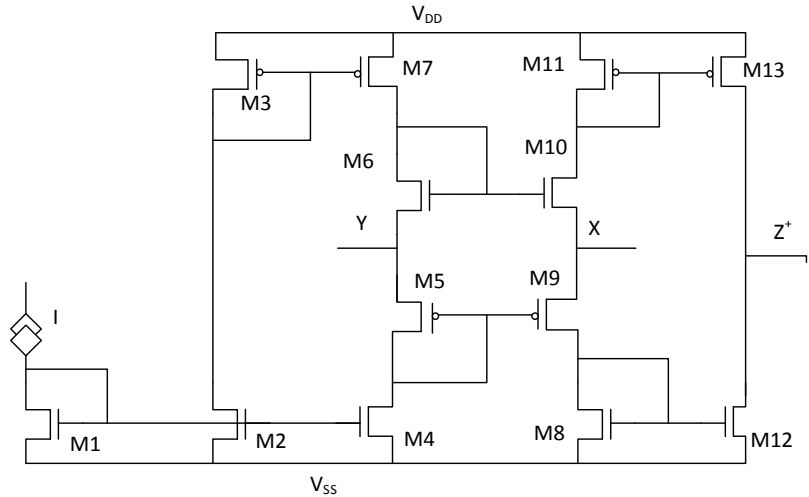

**Figure 5. Translinear circuit used for CCCII+**

## 5. Comprehensive and Compounded Evaluation of Proposed Circuit.

The device analysis results obtained from the simulation of the comparative voltage amplifiers with various circuit topologies and the proposed circuit schemes provided in this work are viz;

**Table. I. Comprehensive Results of the Tunable Voltage Amplifiers.**

| Circuit | No. of CC's | No. of Resistors | Technology | Floating nodes | Average Power Dissipation | Peak Power |
|---|---|---|---|---|---|---|
| Ferri, Guerrini Voltage Amplifier[12] implemented using CCII+ | 2 | 02 | 45nm PTM CMOS | 01 | 1.3154E-05 | 1.3178E-05 |
| Ferri, Guerrini Voltage Amplifier**[12]** implemented using CCII+ | 01 | 02 | 45nm PTM CMOS | nil | 4.9500E-06 | 4.9561E-06 |
| **Ferri, Guerrini Voltage Amplifier implemented using CCCII+** | 02 | 02 | 45nm PTM CMOS | 01 | 2.1233E-04 | 2.2915E-04 |
| **Proposed Voltage Amplifier using CCCII+** | 01 | 02 | 45nm PTM CMOS | nil | 1.1905E-04 | 1.2388E-04 |

Different aspects of amplification are seen on the simulating tool, while tuning the resistors $R_1$ and $R_2$. Such behaviors have been discussed in the Table 2. The tuning of resistances should be done as per the proper guidelined scheme shown under certain cases in the table 2, below inoder to obtain proper amplification.

**Table 2. Scheme for Tuning of the Volatage amplifiers**

| S. no | R1 Verses R2 | Output Behaviour |
|---|---|---|
| CASE-I | R1 greater then R1 | Input signal Amplitude descreses |
| CASE-II | R2 greater than R1 | Ideal Amplifier behaviour |
| CASE-III | R1 equals R2 | Input signal Amplitude descreses |

**6 Output Simulation Results of Amplification of HSPICE Tool**

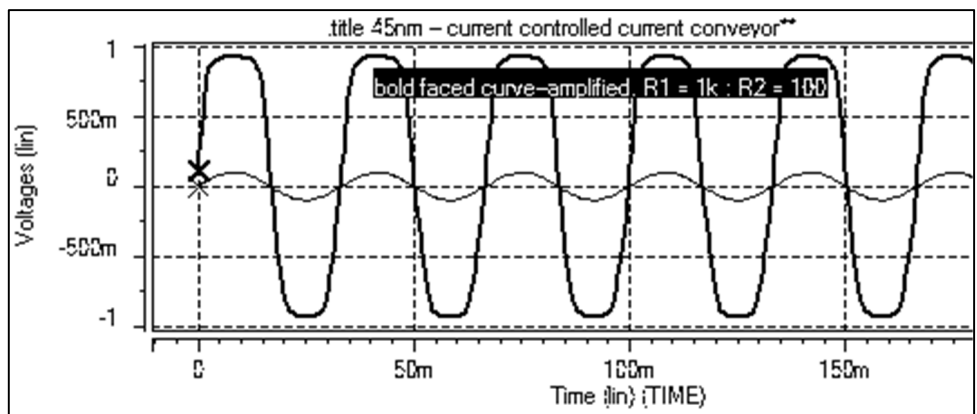

**Figure 6. Voltage Amplification, R1 = 1k and R2 = 100k, $V_{in}$ = 100mV$_{PP}$: $V_{out}$ = 1 V$_{PP}$**

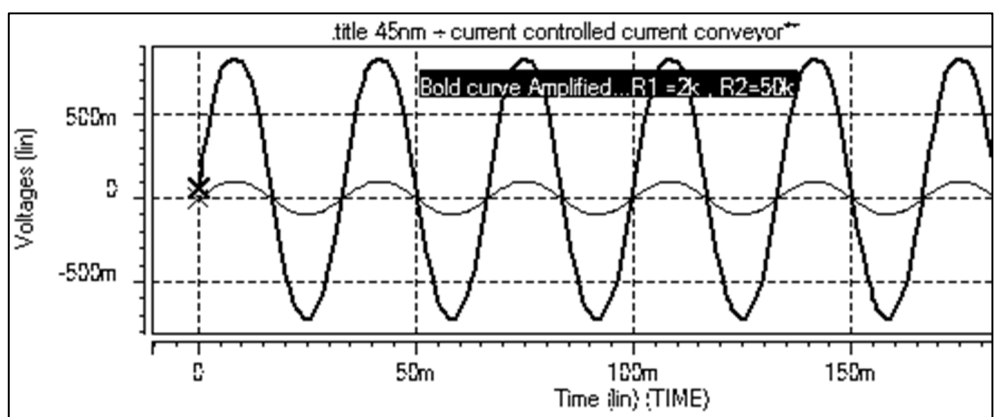

**Figure 6. Voltage Amplification, R1 = 2k and R2 = 50k, $V_{in}$ = 100mV$_{PP}$: $V_{out}$ = 800mV$_{PP}$**

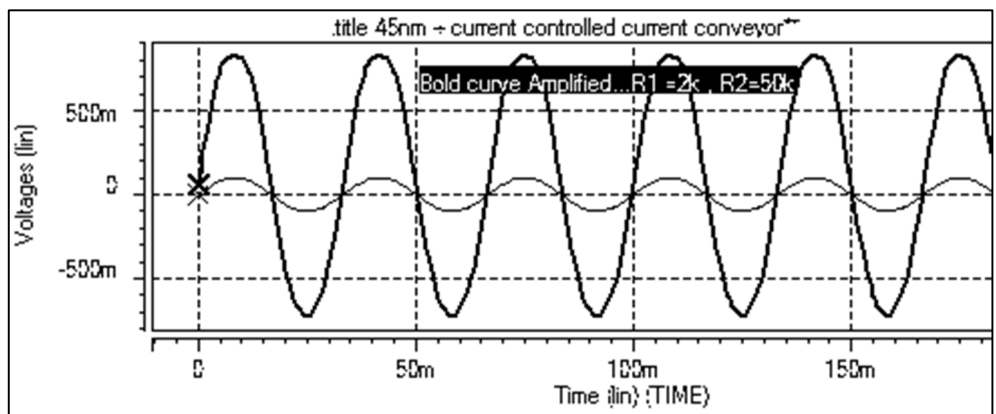

**Figure 6. Voltage Amplification, R1 = 1k and R2 = 100k, $V_{in}$ = 100mV$_{PP}$: $V_{out}$ = 130 V$_{PP}$**

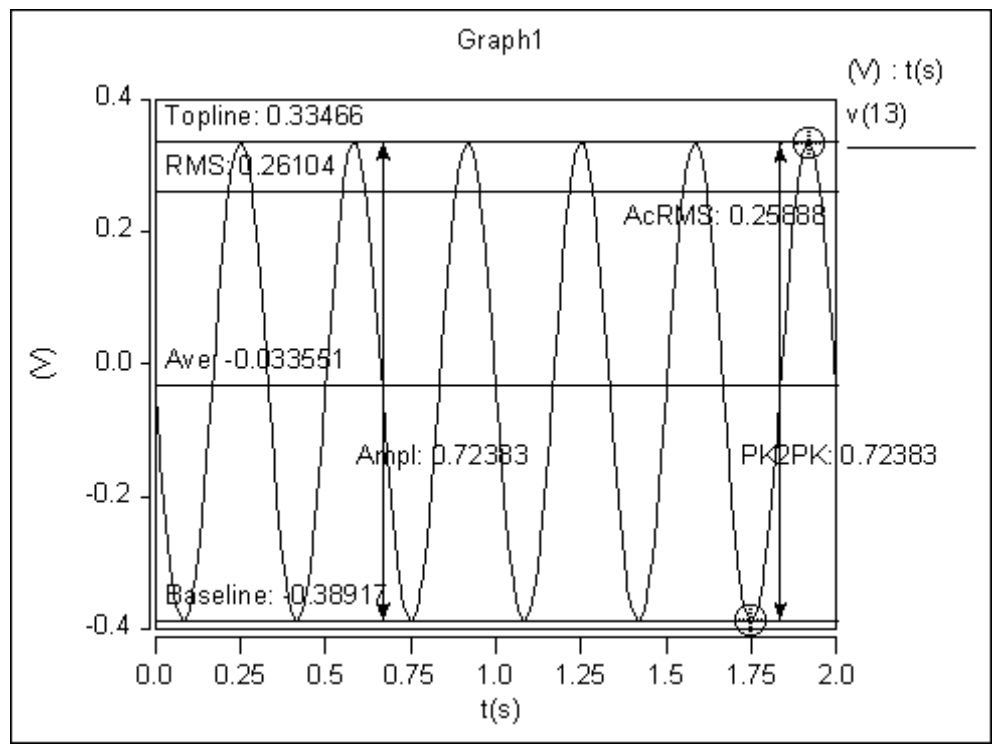

**Figure 7. Calculated RMS values of the O/P waveform with R1=5ohm and R2=10ohm**

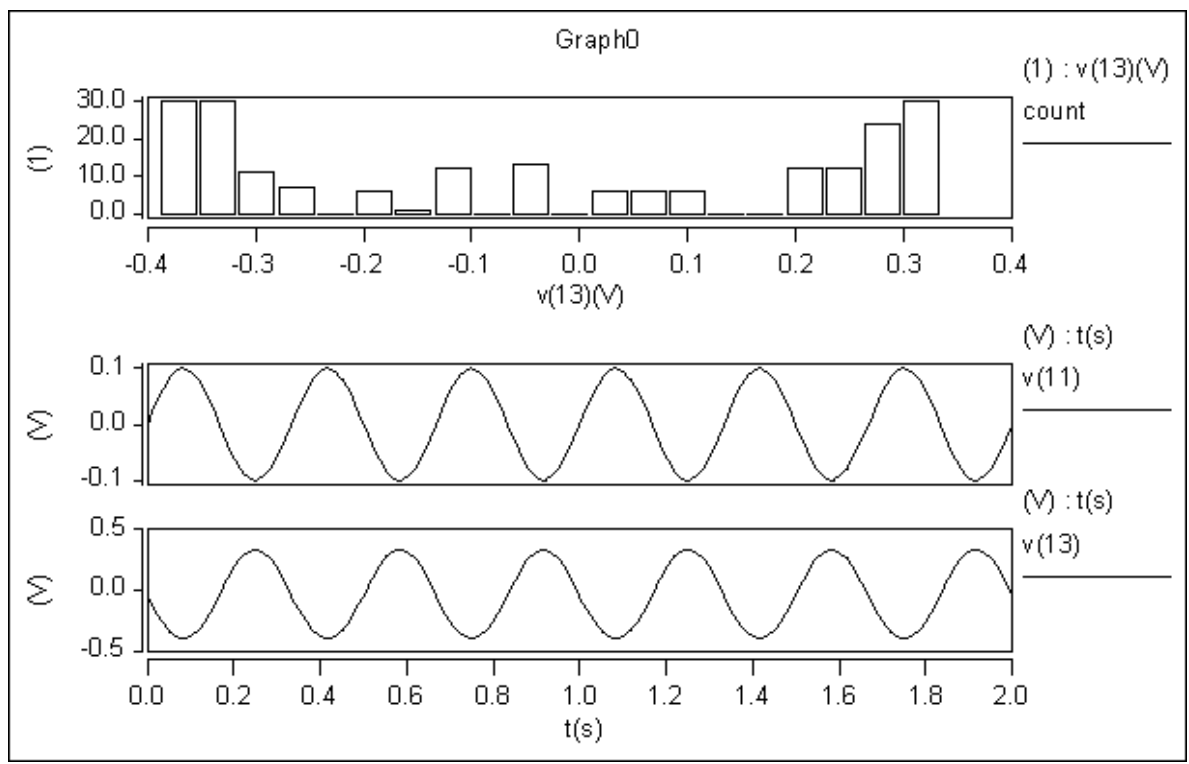

**Figure 8. Histogram of the proposed circuit involving 20 bins. V (11) is the input voltage signal and V (13) is the output voltage signal for particular values of R1 and R2. An increase in the voltage can be seen at the output plot of the waveform cooperating with the proposed theorey irrelation with the Figure 4.**

## 6 CONCLUSION

A novel tunable voltage amplifier employing less hardware is presented in this paper. The proposed voltage amplifier works on the scheme of the Active device Current Conveyor.Proposed Circuit was implemented using the translinear circuit scheme of CCCII+. The simulation results of Proposed circuit were obtained in HSPICE tool in the 45nm Predictive model Mosfet Technology. Proposed design is expected to perform better than already present in the literature interms of the performance parameters Power, PDP delay, Peak Power.In Future , an extended circuit scheme along with the hardware implementation is projected to be presented .

**ACKNOWLEDGEMENT**

The work presented in the paper was carried out at the Signal Processing Lab, Department of Electronics, Islamic University of Science & Technology, Kashmir-192122. The authors would like to thank the staff of the Lab for their kind Support and Patience.

**Biographies of Authors**

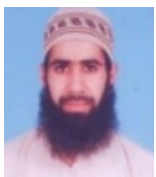

**Umar Mohammad** pursued his bachelor's degree in Electronics and Communication Engineering in 2013 from India. In 2016, he passed Masters in Electronic Circuits and Systems from Integral University Lucknow. Currently, he is pursuing research on Low power Novel MOS devices and also working as Assistant Professor (non-substantive) in the Department of ECE, Islamic University of Science & Technology, Awantipora, and Kashmir, India. He has been active reviewer for various reputed international journals & conferences. He has published many papers in highly indexed journals and conferences. His research interests are Low Power Circuit design, Nano material Circuit design and Novel MOS design.

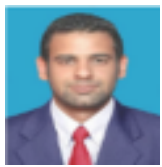

**Mir Aamir Shafi** did his B-Tech in Electronics and Communication Engineering in 2011. I 2014, he passed M-Tech in Power Electronics from Visvesvaraya University, India. Currently, he is working as Assistant Professor in the Department of ECE, Islamic University of Science and Science and Technology, Awanatipora, Kashmir. His research intersests are in Industrual Power Circuit deisgn.